\documentclass[10pt,conference]{IEEEtran}

\usepackage{diagbox}
\usepackage{graphicx}
\usepackage{amsmath}
\usepackage{amsfonts}
\usepackage{algpseudocode}
\usepackage{algorithm}
\usepackage{tikz}
\usepackage{url}
\usepackage{caption}
\usepackage{subcaption}
\usepackage{float}
\usepackage{amsmath,amsfonts,amsthm} 
\usepackage{mathrsfs} 
\usepackage{amssymb}
\usepackage{wasysym}
\usepackage{multirow}
\usepackage{enumerate}
\usepackage{enumitem}
\usetikzlibrary{shapes}
\usetikzlibrary{arrows,automata}
\usetikzlibrary{positioning}
\usetikzlibrary{patterns}
\usetikzlibrary{backgrounds}
\usepackage{array}

\newcommand{\Hi}[1]{$\mathscr{HI}$}
\newcommand{\Lo}[1]{$\mathscr{LO}$}

\algnewcommand{\LeftComment}[1]{\Statex \(\triangleright\) #1}

\newcommand\myeq{\mathrel{\overset{\makebox[0pt]{\mbox{\normalfont\tiny\sffamily def}}}{=}}}
\newcommand{\minisection}[1]{\noindent {\textbf{#1.}}}

\begin{document}

\title{ Utilization Difference Based Partitioned Scheduling of Mixed-Criticality Systems}
\author{
\IEEEauthorblockN{Saravanan Ramanathan, Arvind Easwaran}
\IEEEauthorblockA{Nanyang Technological University, Singapore\\
Email: saravana016@e.ntu.edu.sg, arvinde@ntu.edu.sg}
}
\maketitle
\begin{abstract}
Mixed-Criticality (MC) systems consolidate multiple functionalities with different criticalities onto a single hardware platform. Such systems improve the overall resource utilization while guaranteeing resources to critical tasks. In this paper, we focus on the problem of partitioned multiprocessor MC scheduling, in particular the problem of designing efficient partitioning strategies. We develop two new partitioning strategies based on the principle of evenly distributing the difference between total high-critical utilization and total low-critical utilization for the critical tasks among all processors. By balancing this difference, we are able to reduce the pessimism in uniprocessor MC schedulability tests that are applied on each processor, thus improving overall schedulability. To evaluate the schedulability performance of the proposed strategies, we compare them against existing partitioned algorithms using extensive experiments. We show that the proposed strategies are effective with both dynamic-priority Earliest Deadline First with Virtual Deadlines (EDF-VD) and fixed-priority Adaptive Mixed-Criticality (AMC) algorithms. Specifically, our results show that the proposed strategies improve schedulability by as much as $28.1\%$ and $36.2\%$ for implicit and constrained-deadline task systems respectively.
\end{abstract}

\section{Introduction}
\label{introduction}

Growing complexity in safety-critical real-time systems has led to the concept of consolidating multiple applications with varying criticalities on a common hardware platform. In order to provide guaranteed execution and safe isolation for the critical functions from non-critical ones, the resources are generally reserved using static partitioning. The drawback of this approach is that reserving the resources for the critical functions can potentially lead to under-utilization of the system resources. To overcome this challenge, Vestal~\cite{vestal} proposed the mixed-criticality (MC) model, and since then MC scheduling has received a lot of attention. See~\cite{review} for a very good review on MC scheduling algorithms.

Efficient algorithms that are capable of scheduling a large class of MC systems are highly desired. There are several MC algorithms designed for both uniprocessor and multiprocessor systems. Most of the multiprocessor algorithms are designed based on either global scheduling 
or partitioned scheduling. 
In this work, we focus on partitioned MC scheduling motivated by the fact that 1) they have a better schedulability performance compared to global MC scheduling~\cite{multicoremc}, and 2) safety-critical industries have a preference for them because they are a natural extension of uniprocessor scheduling~\cite{autosar}.

The partitioned MC scheduling problem comprises two main challenges: 1) Statically assigning tasks to processors (\emph{partitioning strategy}), and 2) Scheduling the tasks on each processor using uniprocessor MC scheduling algorithms. The partitioning strategies can be classified into two types: \emph{criticality-aware} partitioning in which tasks of a higher criticality are assigned to processors before tasks of a lower criticality, and \emph{criticality-unaware} partitioning in which no such allocation order exists. The choice of partitioning strategy has a significant impact on the performance of any partitioned scheduling algorithm. Although for conventional (non-MC) systems, first-fit decreasing utilization/density is known to be the best performing strategy, for MC systems there is no known partitioning strategy that performs well in all cases.

In this paper, we propose a new partitioning strategy for dual-criticality MC systems called Utilization Difference based Partitioning (UDP) inspired by the following observation. In dual-criticality systems, there are tasks with two criticality levels, namely \emph{high} and \emph{low}. The high-criticality tasks have two processor utilization values, a \emph{low mode} utilization for execution behaviours in which all tasks are required to meet deadlines, and a \emph{high mode} utilization for execution behaviours in which only the high-criticality tasks are required to meet deadlines. The performance of many uniprocessor MC scheduling algorithms depend on the \emph{sum of difference between these two utilization values} (denoted in short as \emph{utilization difference}) across all the high-criticality tasks assigned to a processor. A smaller utilization difference on a processor generally implies that the MC tasks allocated on that processor is more likely to be schedulable, because the additional demand of high-criticality tasks when the system switches from low to high mode is also small. Therefore, under the UDP strategy we aim to distribute this utilization difference evenly across all processors. 

We propose two partitioning strategies based on UDP; a criticality-aware strategy called \emph{CA-UDP} as well as a criticality-unaware strategy called \emph{CU-UDP}. We use three different scheduling algorithms with both these strategies; Earliest Deadline First with Virtual Deadlines (EDF-VD)~\cite{edfvd2}, Earliest Carry-over Deadline First (ECDF)~\cite{ecdf} and Adaptive Mixed-Criticality (AMC)~\cite{rta}. We chose these algorithms because they cover both dynamic and fixed-priority schemes, and their schedulability tests are such that the pessimism in those tests can be reduced by a smaller utilization difference. Note that the EDF-VD test also has an optimal speed-up bound\footnote{Speed-up bound denotes the smallest increase in processor-speed necessary to schedule all feasible MC task systems.} of $4/3$ for implicit-deadline\footnote{Systems in which relative deadlines of tasks are equal to their minimum release separation time or period.} MC task systems, implying that the resulting partitioned MC scheduling algorithms also have a speed-up bound of $8/3$ (see Section~\ref{sec:model} for a discussion on this speed-up bound).

We performed extensive experimental evaluation of the proposed partitioned scheduling algorithms and compared them with existing algorithms. For implicit-deadline task systems, the schedulability improvement using our algorithms is as much as $28.1\%$
with EDF-VD when compared to an existing algorithm with a known speed-up bound, and $15.7\%$ with ECDF and $9.5\%$ with AMC when compared to existing algorithms without a known speed-up bound. For constrained-deadline\footnote{Systems in which relative deadlines of tasks are no more than their minimum release separation time or period.} task systems, the schedulability improvement using our algorithms is as much as $29.7\%$ with AMC and $36.2\%$ with ECDF when compared to existing algorithms. We also observed that the performance of our algorithms improve significantly with increasing number of processors, indicating their scalability. Among the two partitioning strategies, \emph{CA-UDP} and \emph{CU-UDP}, \emph{CU-UDP} has a better performance overall because it allocates high utilization low-criticality tasks earlier, and hence is more likely to find a feasible allocation for such heavy utilization tasks.

\textbf{Related Work.} There have been few studies on partitioned MC scheduling algorithms. Kelly et al.~\cite{kelly} introduced the fixed-priority, partitioned multiprocessor MC scheduling and showed that decreasing criticality and first-fit partitioning performs well in comparison to decreasing utilization and worst-fit partitioning. Lakshmanan et al.~\cite{cop} presented a criticality-aware hybrid partitioning strategy with decreasing utilization. Baruah et al.~\cite{multicoremc} presented a partitioned scheduling algorithm based on EDF-VD and proved that the algorithm has a speedup of $8/3$. They also showed that partitioned scheduling performs better than the global variant in terms of schedulability. Rodriguez et al.~\cite{rodriguez} evaluated different partitioning strategies under EDF-VD and showed that criticality-aware partitioning performs better than criticality-unaware partitioning. Guan et al.~\cite{mpvd} presented a criticality-aware partitioning strategy with worst-fit allocation for high-criticality tasks and first-fit allocation for low-criticality tasks, additionally also giving preference to heavy utilization low-criticality tasks. They applied this partitioning strategy to a demand bound function based test and virtual deadline based algorithm presented in~\cite{ey}. None of the above studies considered partitioning strategies based on evenly distributing the utilization difference of high-criticality tasks among all processors.
\section{System Model and Scheduling Algorithms}
\label{sec:model}

We consider a dual-criticality (namely LC for low-criticality and HC for high-criticality) sporadic task system $\tau$ comprising $n$ tasks scheduled on a multiprocessor platform with $m$ cores. Each task $\tau_i$ is defined by a tuple ($T_i$,$\chi_i$,$C_i^L$,$C_i^H$,$D_i$), where $T_i$ $\in\mathbb{R}^+$ denotes the minimum release separation time, $\chi_i \in\{LC,HC\}$ is the criticality level of the task, $C_i^L$ and $C_i^H$ are the LC and HC execution requirements respectively (we assume $C_i^L$ $\leq$ $C_i^H$ for HC tasks), and $D_i \in\mathbb{R}^+$ is the relative deadline of the task; $D_i = T_i$ for all $i$ in the case of an implicit-deadline task system and $D_i \leq T_i$ for all $i$ in the case of a constrained-deadline task system.

LC task set $\tau_L$ and HC task set $\tau_H$ are defined as $\tau_L\myeq\{ \tau_i\in\tau\mid \chi_i=LC\}$ and $\tau_H\myeq\{\tau_i\in\tau\mid \chi_i=HC\}$ respectively. Also, the LC and HC utilization of a task $\tau_i$ is $u_i^L\myeq C_i^L/T_i$ and $u_i^H\myeq C_i^H/T_i$ respectively. Normalized system-level utilizations are defined as $U_L^L\myeq\sum_{\tau_i\in\tau_L} u_i^L/m$, $U_H^L\myeq\sum_{\tau_i\in\tau_H} u_i^L/m$ and $U_H^H\myeq\sum_{\tau_i\in\tau_H} u_i^H/m$. 

The system is said to be either in the LC mode or in the HC mode. If all tasks $\tau_i\in\tau$ signal completion before exceeding their LC execution requirement, the system is said to be in the LC mode and all task deadlines are required to be met in this mode. If any HC task $\tau_i\in\tau_H$ executes beyond its LC execution requirement, but signals completion before exceeding its HC execution requirement, the system is said to be in the HC mode. Mode switch instant is defined as the first time instant in a busy interval when the system mode changes from LC to HC. No LC task deadlines are required to be met after mode switch, and hence several MC algorithms immediately discard all LC tasks.
 
We now briefly describe the uniprocessor MC scheduling algorithms and tests used in our experiments. We consider three of them: fixed-priority Adaptive Mixed-Criticality (AMC)~\cite{rta}, dynamic-priority Earliest Deadline First with Virtual Deadlines (EDF-VD)~\cite{edfvd2}, and dynamic-priority Earliest Carry-over Deadline First (ECDF)~\cite{ecdf}. This choice of algorithms is motivated by two facts: 1) it enables us to show that the proposed partitioning strategies work well with both dynamic as well as fixed-priority algorithms, and 2) all these algorithms have schedulability tests whose pessimism can be reduced by reducing the difference between total HC and LC utilizations of HC tasks allocated on a processor; therefore these tests are well suited to be used with the proposed strategies.       

Under AMC, each task has a fixed-priority and all LC tasks are immediately dropped upon a mode switch. We use a previously developed response time analysis based uniprocessor schedulability test for this algorithm (AMC-max test in~\cite{rta}). Since a worst-case mode switch instant is unknown, this test considers all possible mode switch instants until the low mode response time of the task.

Under EDF-VD, each HC task is assigned a virtual deadline to be used in the LC mode that is no larger than the actual deadline. These virtual deadlines are assigned using a single scaling factor so that the slack in the low mode is uniformly distributed among all HC tasks. Tasks are then scheduled using EDF in both modes, and LC tasks are immediately dropped upon a mode switch. For implicit-deadline task systems, an utilization based uniprocessor schedulability test for EDF-VD with an optimal speed-up bound of $4/3$ has been developed (Theorems~$1$ and~$2$ in~\cite{edfvd2}). This test has also been combined with a simple first-fit partitioning strategy to derive a partitioned scheduling algorithm with a speed-up bound of $8/3$~\cite{multicoremc}. In fact, this work shows that any partitioning strategy which considers all processors for allocation of a task before declaring failure has the same speed-up bound of $8/3$ when used with this test (Theorem~$9$ in \cite{multicoremc}). \emph{Since our proposed strategies have this property, by combining the strategies with this EDF-VD test, the resulting partitioned algorithms also have a speed-up bound of $8/3$ for implicit-deadline task systems}.   

Similar to EDF-VD, ECDF is also based on assigning virtual deadlines to HC tasks in the LC mode, and uses EDF for scheduling. But, unlike EDF-VD, it uses a demand bound function based test and a greedy deadline assignment strategy (Theorem~$2$ in~\cite{ecdf}). This test can be used for both implicit as well as constrained-deadline task systems, but it does not have a known speed-up bound.

\minisection{Partitioned versus global scheduling} There is a fundamental difference in the behaviours of partitioned versus global MC scheduling algorithms when a mode switch is triggered. Under global scheduling, when a HC task on a processor triggers a mode switch, it is reflected instantaneously on the other processors, meaning all LC tasks are discarded thereafter. Whereas, under partitioned scheduling, the mode switch is restricted only to that particular processor. Essentially, the tasks executing on other processors, including LC tasks, are unaffected by the mode switch. This isolation between processors is feasible under partitioned scheduling because the MC schedulability tests are applied independently on each processor. It is an important property because it also reduces the impact on LC tasks, and this could be one of the reasons why partitioned MC scheduling may find preference in safety-critical industries.

\section{Utilization Difference Based Partitioning (UDP) Strategies}
\label{design}

In this section, we present two partitioning strategies based on evenly distributing the utilization difference among all processors; a criticality-aware strategy called CA-UDP and a criticality-unaware strategy called CU-UDP. The guiding principle behind both these strategies is the same, which is to reduce the maximum gap between the total HC utilization and total LC utilization of the HC tasks allocated on each processor, i.e., $\max \{ U_H^H(\phi_k)-U_H^L(\phi_k) | 1 \leq k \leq m \}$.

We use $\phi_k$ to denote the $k^{th}$ processing core. At any point during the partitioning strategy, we use $\tau(\phi_k)$ to denote the set of tasks assigned to processor $\phi_k$ until that point. Then, the system-level utilizations for processor $\phi_k$ at that point can be defined as follows: $U_L^L(\phi_k)\myeq\sum_{\tau_i\in\tau(\phi_k)\wedge\chi_i=LC} u_i^L$, $U_H^L(\phi_k)\myeq\sum_{\tau_i\in\tau(\phi_k)\wedge\chi_i=HC} u_i^L$ and $U_H^H(\phi_k)\myeq\sum_{\tau_i\in\tau(\phi_k)\wedge\chi_i=HC} u_i^H$.

Algorithm~\ref{algo:1} presents a detailed pseudocode for CA-UDP. Under this strategy, all HC tasks ($\tau_H$) are assigned to processors before assigning any of the LC tasks. To successfully assign tasks having very high utilization values, we sort the tasks in decreasing order of utilization values at their respective criticality levels, i.e., tasks in $\tau_H$ are sorted based on $u_i^H$, whereas tasks in $\tau_L$ are sorted based on $u_i^L$. For tasks in $\tau_H$, CA-UDP uses a worst-fit allocation strategy based on the parameter $U_H^H(\phi_k)-U_H^L(\phi_k)$, i.e., processors are considered in increasing order of $U_H^H(\phi_k)-U_H^L(\phi_k)$. This strategy ensures an even distribution of the utilization difference among all processors. For tasks in $\tau_L$, CA-UDP uses a simple first-fit allocation strategy. Before assigning any task to a processor, we evaluate the feasibility of this allocation using one of the schedulability tests mentioned in Section~\ref{sec:model} depending on the chosen scheduling algorithm. If the test fails on a processor, then the next processor in the fitting order is considered. If the test fails on all processors, then the partitioning is declared a failure.

\begin{algorithm}
\caption{CA-UDP}
\label{algo:1}
\begin{algorithmic}[1]
\LeftComment {Partitioning $\tau_H$}
\State Sort $\tau_H$ in decreasing order of $u_i^H$.
\For{$j := 1$ to length($\tau_H$)}
	\State Sort $\phi_k$ in increasing order of $U_H^H(\phi_k)-U_H^L(\phi_k)$.
	\For{$k := 1$ to $m$}
		\If{$\tau(\phi_k) \cup \tau_i$ is schedulable}
			\State $\tau(\phi_k) = \tau(\phi_k) \cup \tau_i$ and break.
		\EndIf
	\EndFor
	\If{$\tau_i$ could not be allocated}
		\State Return partitioning failed.
	\EndIf
	\EndFor
\LeftComment {Partitioning $\tau_L$}
\State Sort $\tau_L$ in decreasing order of $u_i^L$.
\For{$j := 1$ to length($\tau_L$)}
	\For{$k := 1$ to $m$}
		\If{$\tau(\phi_k) \cup \tau_i$ is schedulable}
			\State $\tau(\phi_k) = \tau(\phi_k) \cup \tau_i$ and break.
		\EndIf
	\EndFor
	\If{$\tau_i$ could not be allocated}
		\State Return partitioning failed.
	\EndIf
\EndFor
\State Return $\tau(\phi_k)$, $\forall k\in[1,\ldots,m]$.
\end{algorithmic}
\end{algorithm}

\begin{figure}[h]
	\centering
	\includegraphics[width=0.49\textwidth,height=3.5cm]{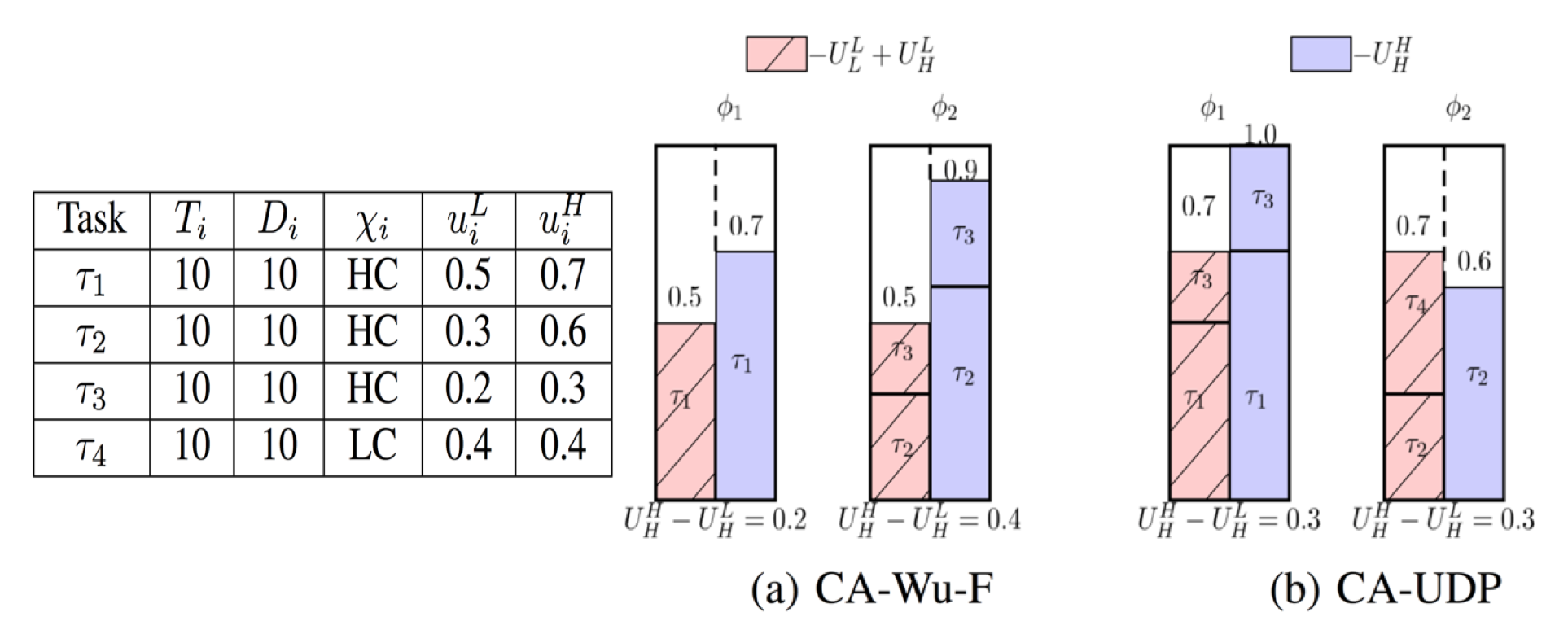}
	\caption{Comparison of CA-UDP and CA-Wu-F}\label{fig:example}
	\vspace{-0.5cm}
\end{figure}

\minisection{Example} Consider an example task set as shown in Figure~\ref{fig:example} to be scheduled on $2$ processors ($\phi_1$ and $\phi_2$) using partitioned EDF-VD algorithm. To illustrate the benefit of CA-UDP, we compare it with another criticality-aware partitioning strategy that uses worst-fit allocation based on total HC utilization alone for HC tasks and an identical strategy as CA-UDP for LC allocations (denoted as CA-Wu-F). Figure~\ref{fig:example} shows the task allocation under both these strategies when applied to EDF-VD. Under CA-Wu-F, task $\tau_1$ is assigned to $\phi_1$ and tasks $\tau_2$ and $\tau_3$ are assigned to $\phi_2$. Although task $\tau_4$ can fit on either $\phi_1$ or $\phi_2$ based on its utilization, it fails to get allocated on any of the processors because the EDF-VD schedulability test $\big(U_L^L(\phi_k) \leq (1-U_H^H(\phi_k))/(1-(U_H^H(\phi_k)-U_H^L(\phi_k)))\big)$ fails. Whereas, under CA-UDP, tasks $\tau_1$ and $\tau_3$ are allocated on one processor and task $\tau_2$ is allocated on the other processor in order to balance the utilization difference. Then, task $\tau_4$ can be successfully allocated on the processor with $\tau_2$. Thus, by balancing the utilization difference, we are able to provide more processor choices for LC tasks, which consequently improves schedulability.
%
%
One of the challenges with a criticality-aware partitioning strategy such as CA-UDP is that it often fails to allocate LC tasks with very high utilization values, because these tasks are considered after all the HC tasks have been allocated. To address this problem, a simple approach is to consider a criticality-unaware partitioning strategy in which all tasks (HC and LC) are allocated in the decreasing order of their utilization values at their respective criticality levels. Based on this principle of criticality-unaware allocation, we propose another UDP strategy called CU-UDP. Under this strategy, all tasks in $\tau$ are collectively sorted in decreasing order of utilization; $u_i^H$ is used as utilization for tasks in $\tau_H$ and $u_i^L$ is used as utilization for tasks in $\tau_L$. The rest of the strategy is identical to CA-UDP; a worst-fit allocation strategy based on the parameter $U_H^H(\phi_k)-U_H^L(\phi_k)$ for HC tasks, and a simple first-fit allocation strategy for LC tasks.

\begin{figure}[h]
	\centering
	\includegraphics[width=0.49\textwidth,height=3.5cm]{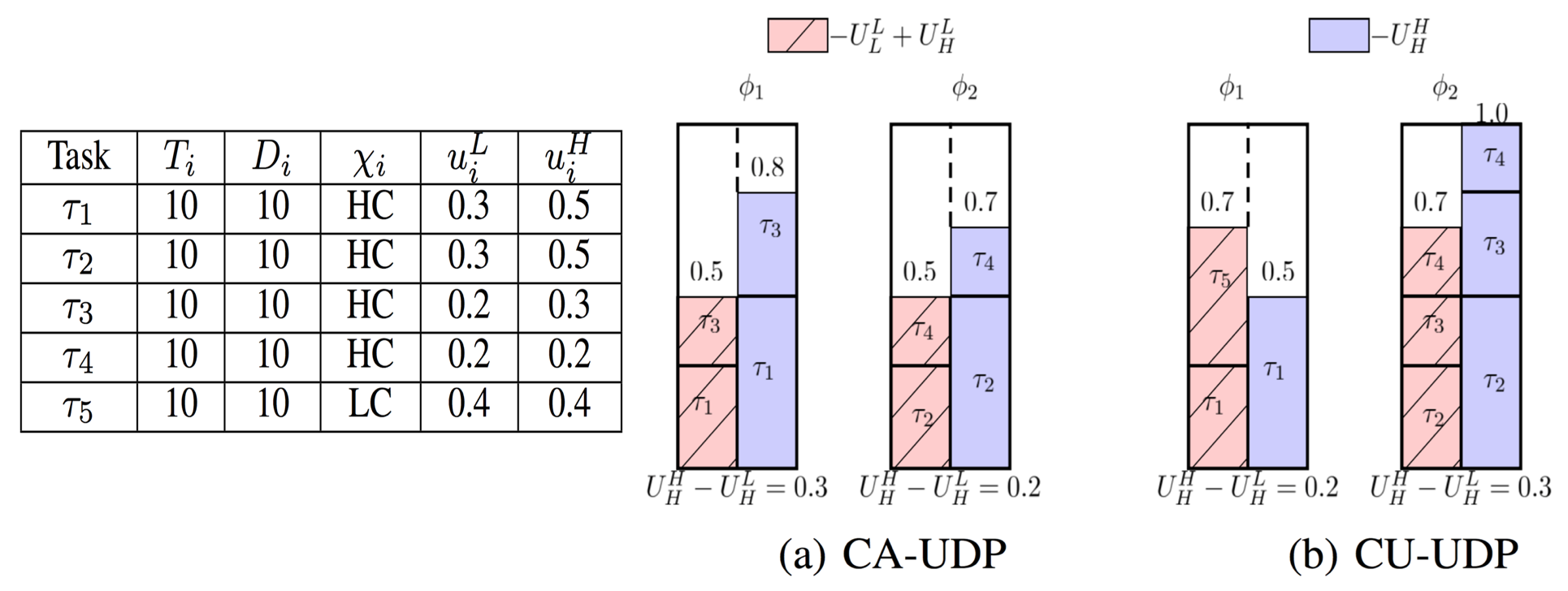}
	\caption{Comparison of CA-UDP and CU-UDP}\label{fig:example2}
	\vspace{-0.1cm}
\end{figure}

\minisection{Example} Figure~\ref{fig:example2} shows an example task set to be scheduled on $2$ processors using EDF-VD. For this example, CU-UDP gives a successful partition while CA-UDP fails as shown in Figure~\ref{fig:example2}. Under CA-UDP, tasks $\tau_1$ and $\tau_3$ are allocated to $\phi_1$ and tasks $\tau_2$ and $\tau_4$ are allocated to $\phi_2$. As a result, the EDF-VD schedulability test fails on both processors when allocating $\tau_5$. In the case of CU-UDP, tasks $\tau_1$ and $\tau_2$ are allocated to $\phi_1$ and $\phi_2$ respectively. Also, due to criticality-unaware partitioning, task $\tau_5$ is allocated to $\phi_1$ before tasks $\tau_3$ and $\tau_4$. Then, tasks $\tau_3$ and $\tau_4$ are successfully allocated to $\phi_2$. In this example, CU-UDP was able to prioritize allocation for a heavy utilization LC task $\tau_5$ while still balancing the utilization difference, and hence gave a successful partition.
\section{Evaluation}
\label{experiments}

In this section, we evaluate the proposed partitioning strategies by combining them with three uniprocessor MC scheduling algorithms as discussed in Section~\ref{sec:model}, EDF-VD, ECDF and AMC. Specifically, we compare their schedulability against the following existing partitioned MC algorithms: 
\begin{itemize}
\item CA(nosort)-F-F-EDF-VD~\cite{multicoremc}, which is the only partitioned scheduling algorithm with a known speed-up bound. It uses criticality-aware partitioning without sorting the tasks, first-fit allocation strategy for both HC as well as LC tasks and employs EDF-VD algorithm.
\item ECA-Wu-F-EY~\cite{mpvd}, which is an enhanced critical-aware partitioning strategy with worst-fit allocation based on HC utilization alone for HC tasks and a first-fit allocation for LC tasks. The enhancement is that preference is given to heavy utilization LC tasks over HC tasks, while the rest of the allocation uses critical-aware partitioning. The tasks are sorted based on the utilization values at their respective criticality levels before being allocated. ECA-Wu-F-EY employs a virtual deadline based uniprocessor MC scheduling algorithm called EY that is identical to EDF-VD and ECDF~\cite{ey}, except that the schedulability test is based on demand bound functions as in ECDF but it is relatively less efficient in terms of schedulability. 
\item CA-F-F-EY~\cite{rodriguez}, is also a criticality-aware partitioning strategy with a simple first-fit allocation for both HC and LC tasks. The tasks are sorted based on utilization values at their respective criticality levels before being allocated, and scheduled using EY algorithm. It has been previously shown that this algorithm dominates all other existing criticality-aware partitioning strategies in experiments.
\end{itemize}

\begin{figure*}
\begin{subfigure}[b]{0.32\linewidth}
\includegraphics[width=\textwidth,height=4cm]{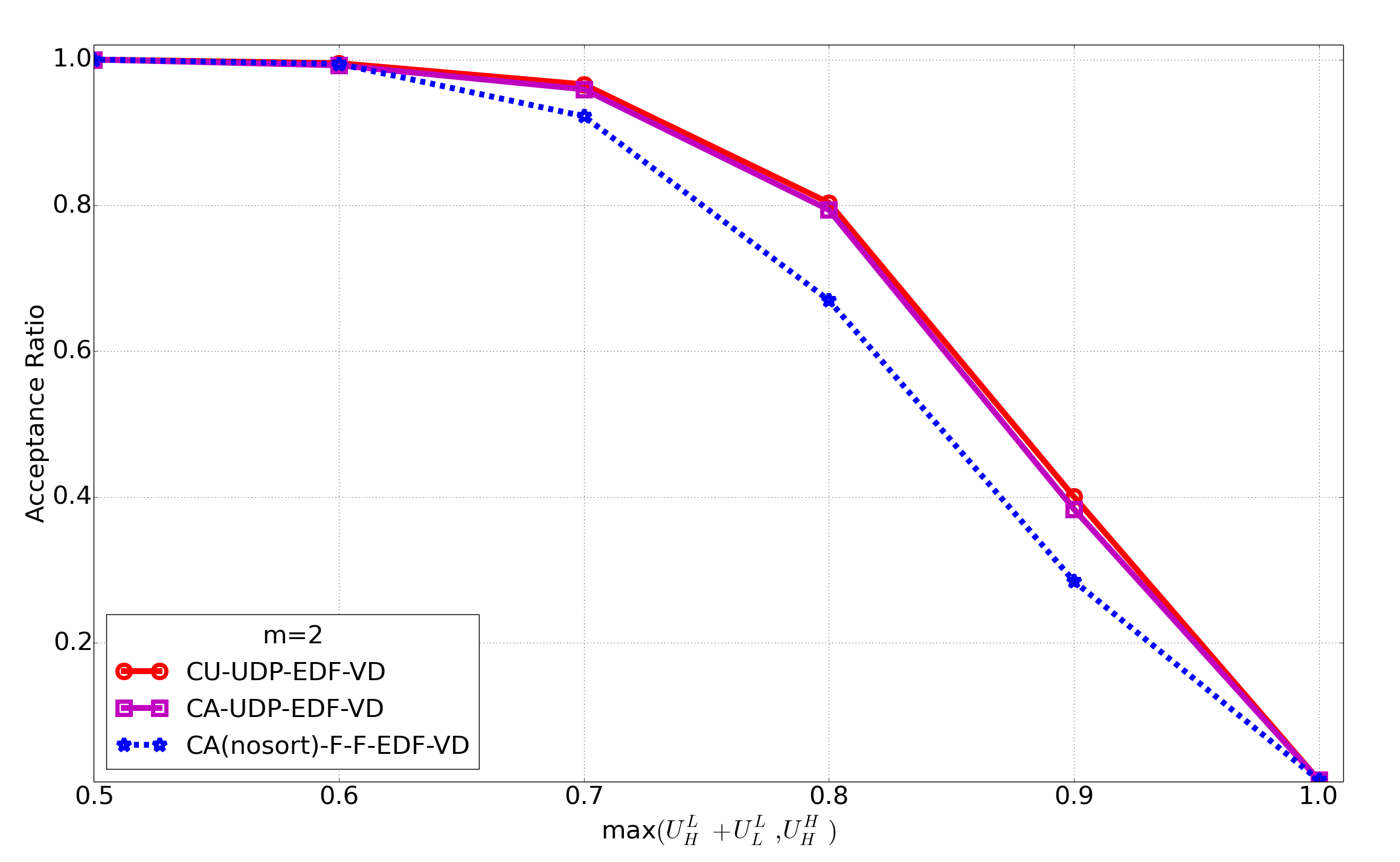}
\caption{EDF-VD ($m=2$)}
\label{fig:2a}
\end{subfigure}
\begin{subfigure}[b]{0.32\linewidth}
\includegraphics[width=\textwidth,height=4cm]{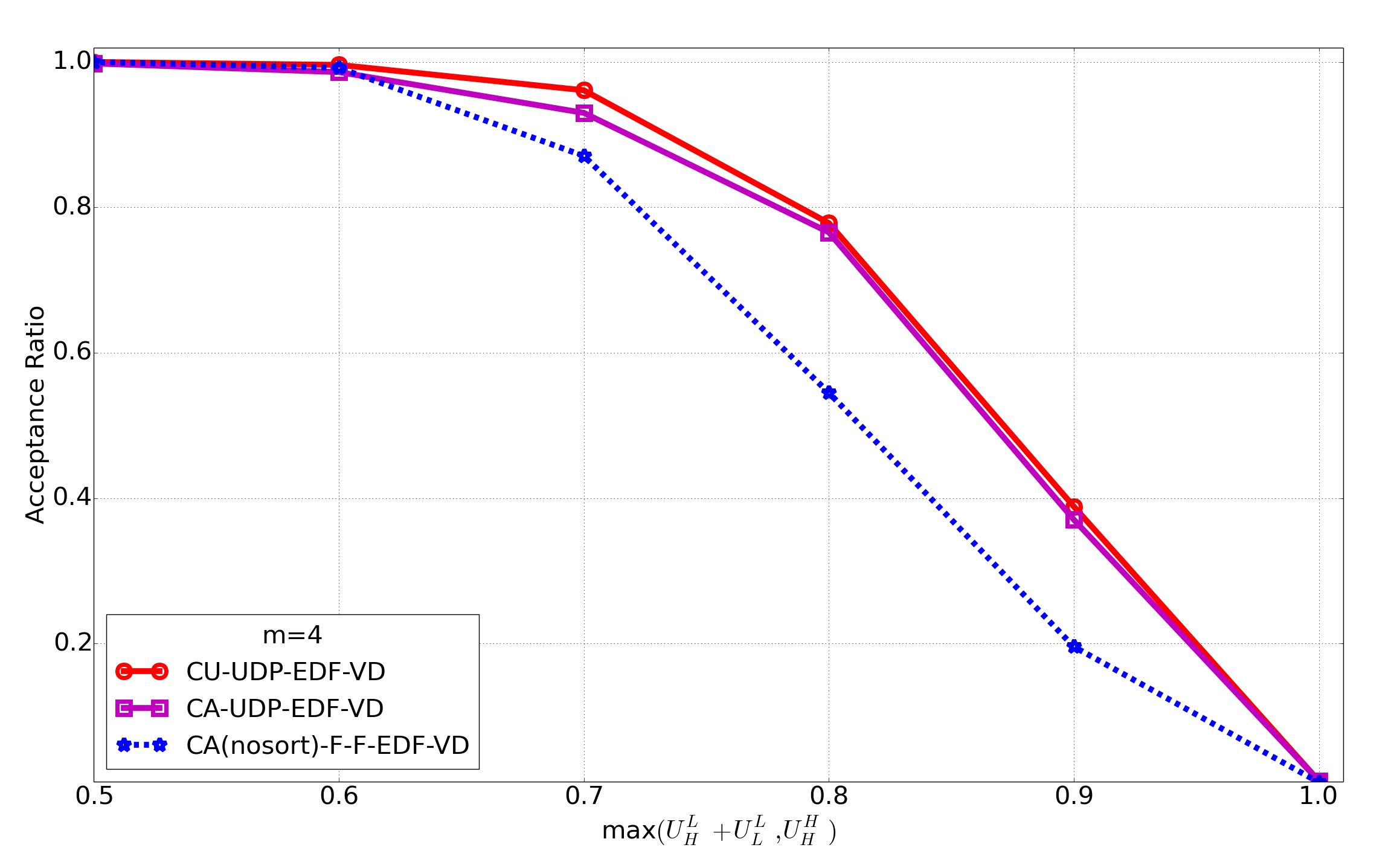}
\caption{EDF-VD ($m=4$)}
\label{fig:4a}
\end{subfigure}
\begin{subfigure}[b]{0.32\linewidth}
\includegraphics[width=\textwidth,height=4cm]{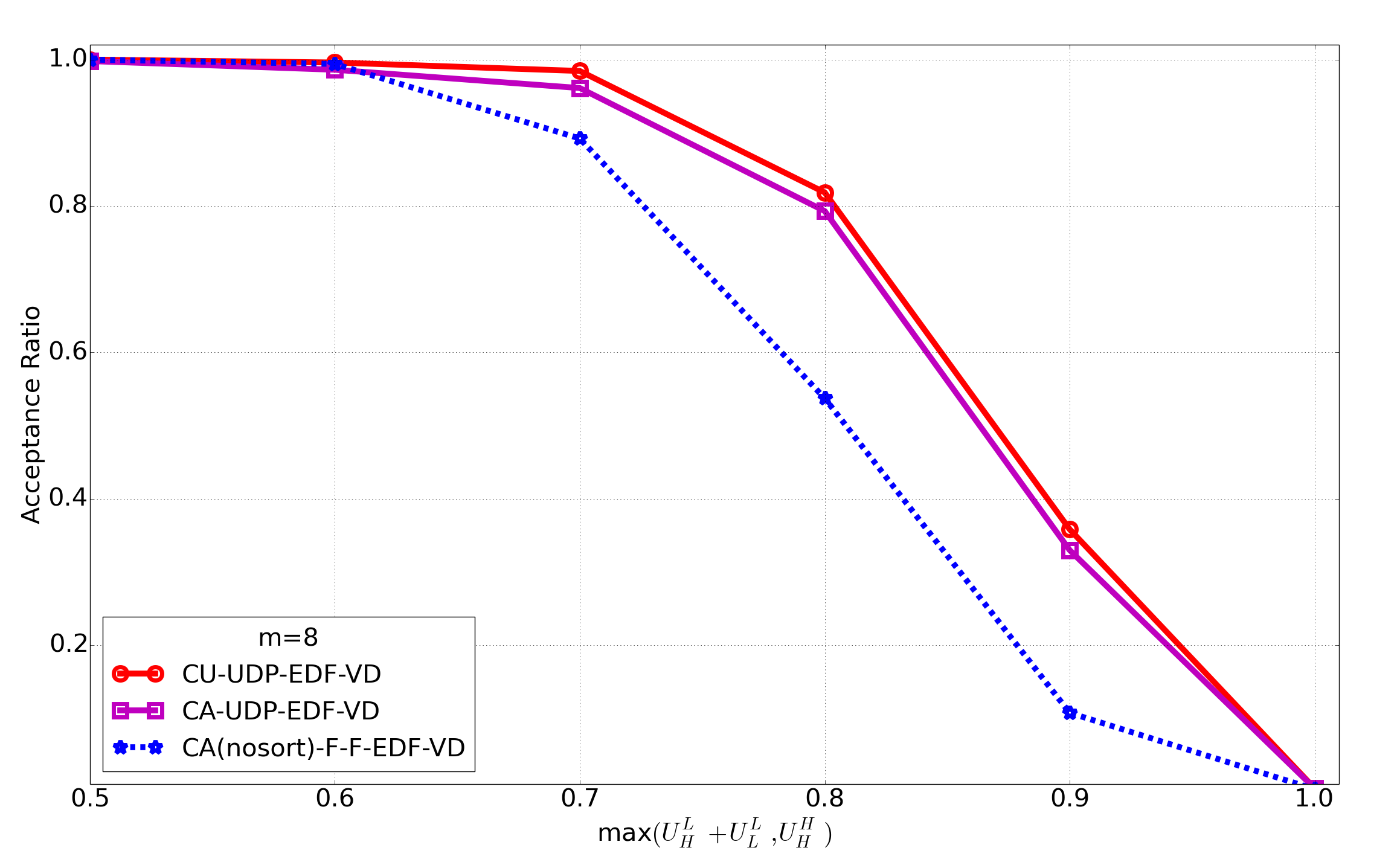}
\caption{EDF-VD ($m=8$)}
\label{fig:8a}
\end{subfigure}
\caption{Comparison of Acceptance Ratio (implicit-deadline with speed-up bound)}\label{schedulability_implicit_speed_up}
\vspace{-0.1cm}
\end{figure*}
\begin{figure*}
\begin{subfigure}[b]{0.32\linewidth}
\includegraphics[width=\textwidth,height=4cm]{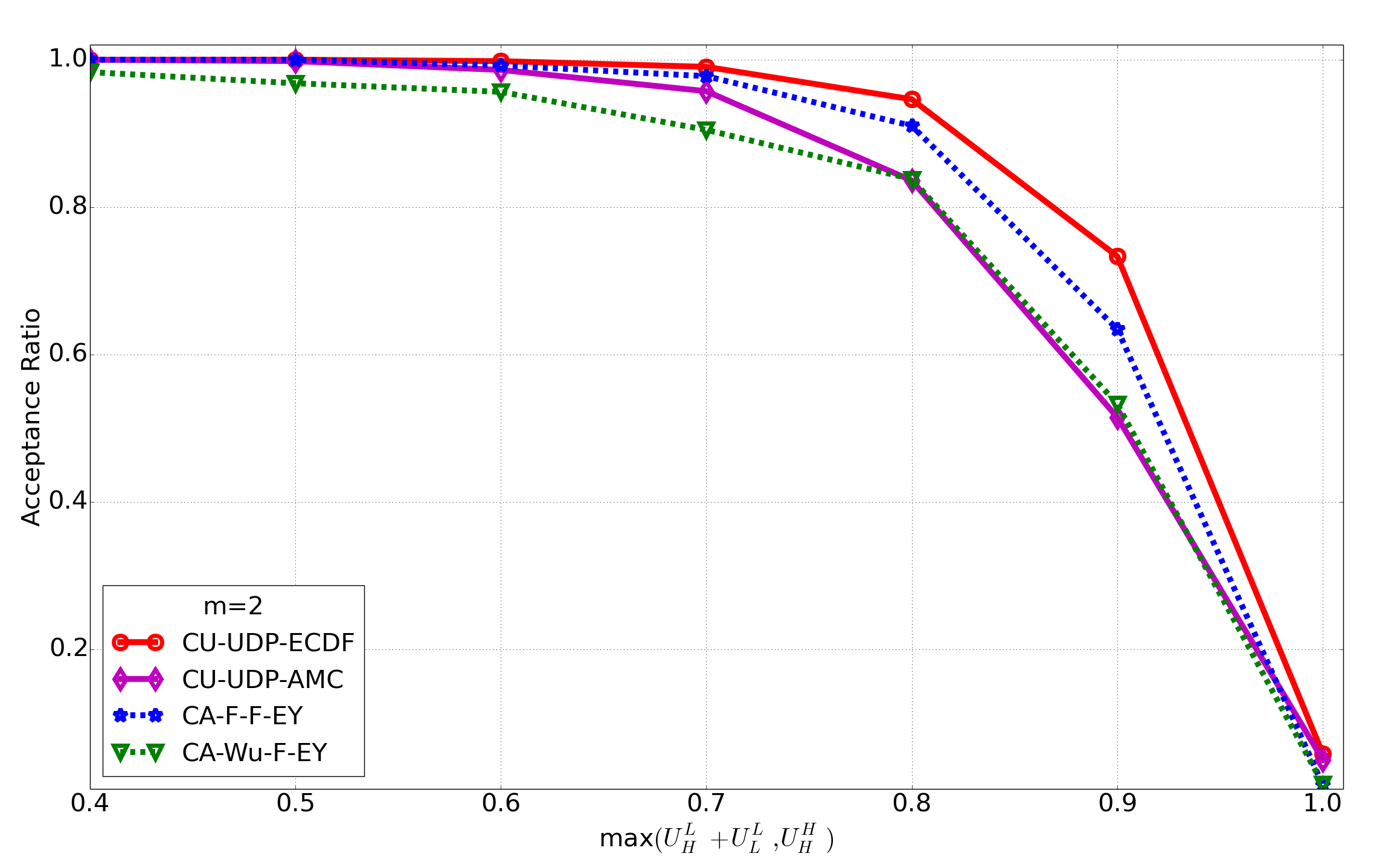}
\caption{$m=2$}
\label{fig:2b}
\end{subfigure}
\begin{subfigure}[b]{0.32\linewidth}
\includegraphics[width=\textwidth,height=4cm]{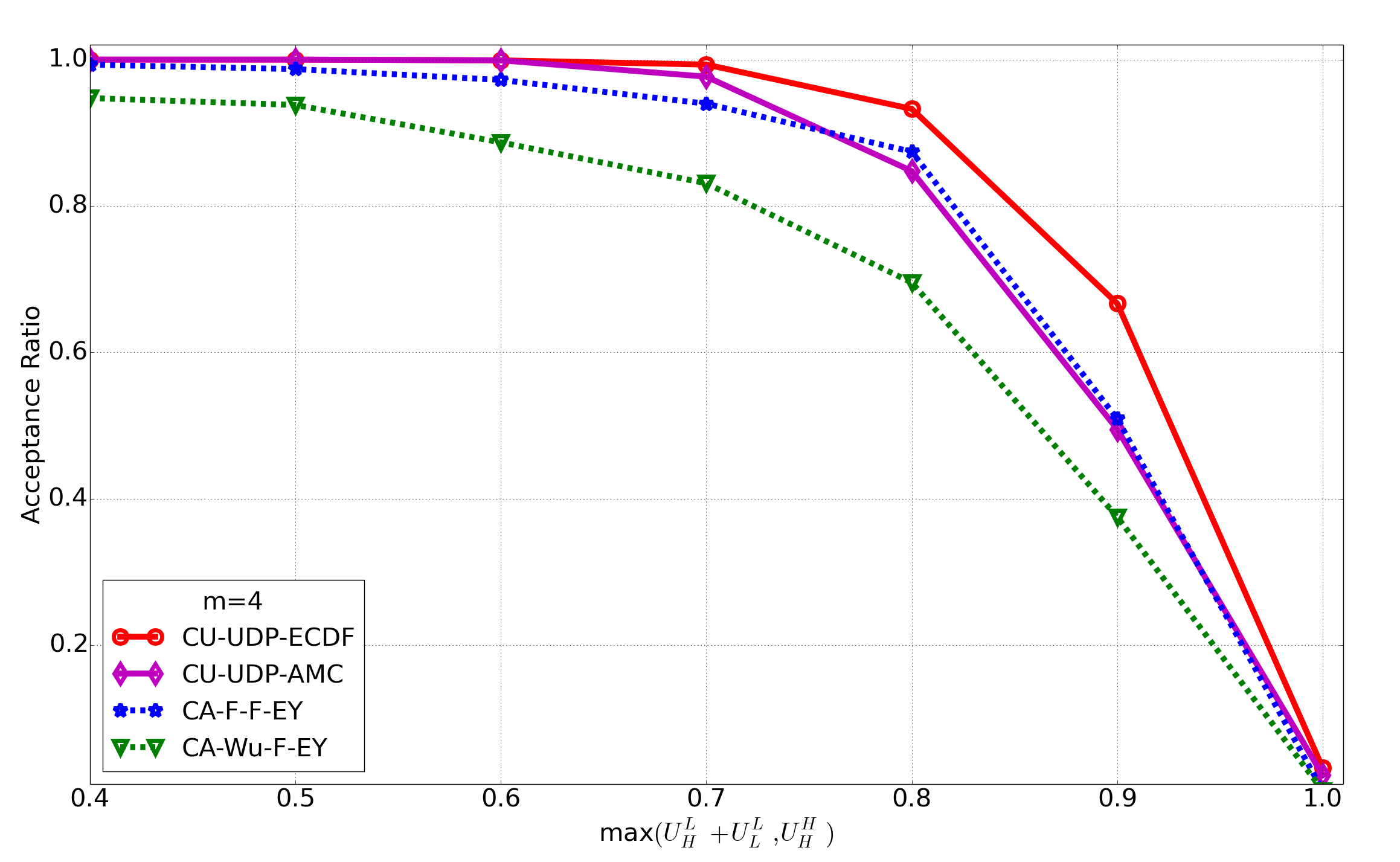}
\caption{$m=4$}
\label{fig:4b}
\end{subfigure}
\begin{subfigure}[b]{0.32\linewidth}
\includegraphics[width=\textwidth,height=4cm]{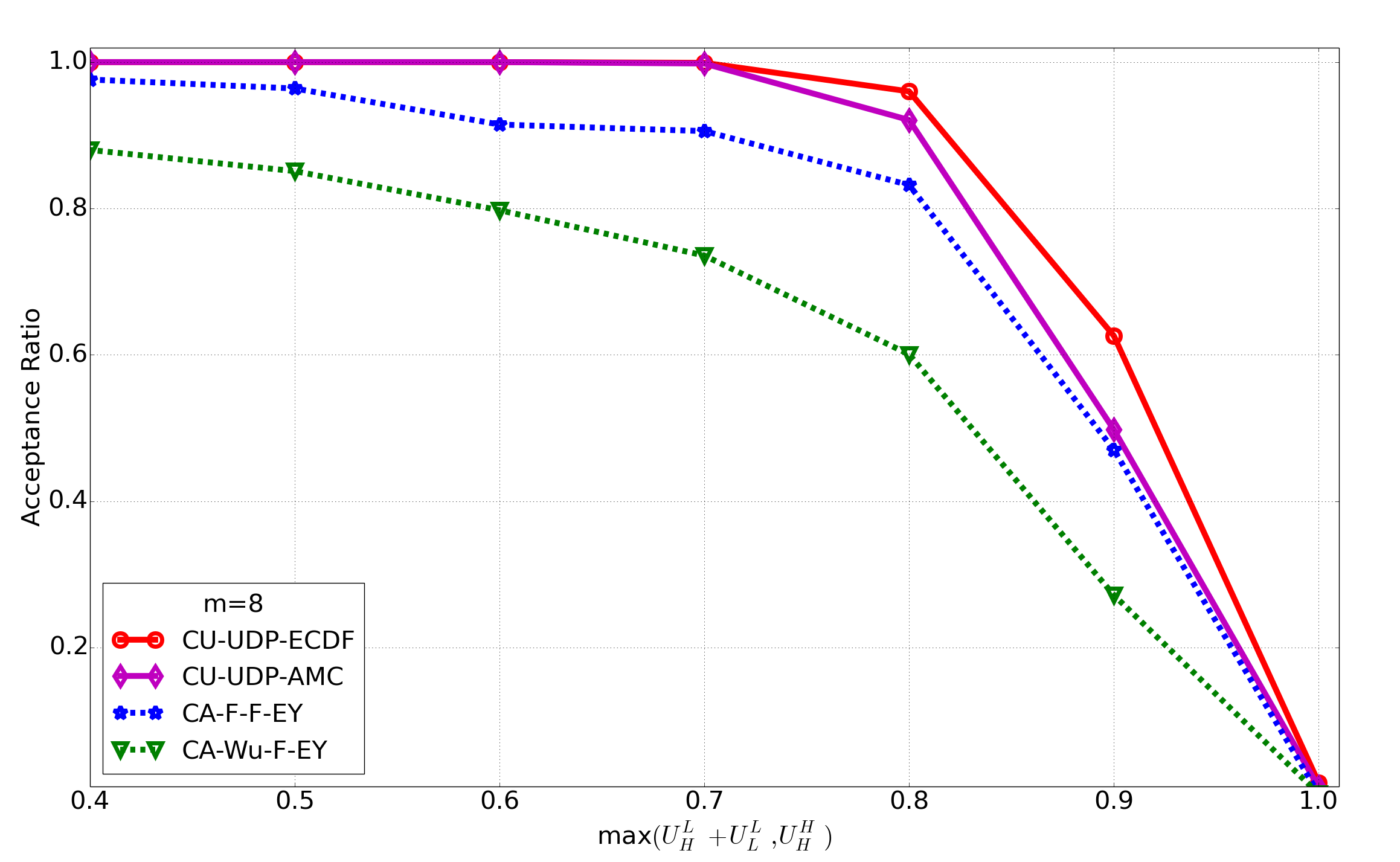}
\caption{$m=8$}
\label{fig:8b}
\end{subfigure}
\caption{Comparison of Acceptance Ratio (implicit-deadline w/o speed-up bound)}\label{schedulability_implicit}
\vspace{-0.2cm}
\end{figure*}
\begin{figure*}
\centering
\begin{subfigure}[b]{0.32\linewidth}
\includegraphics[width=\textwidth,height=4cm]{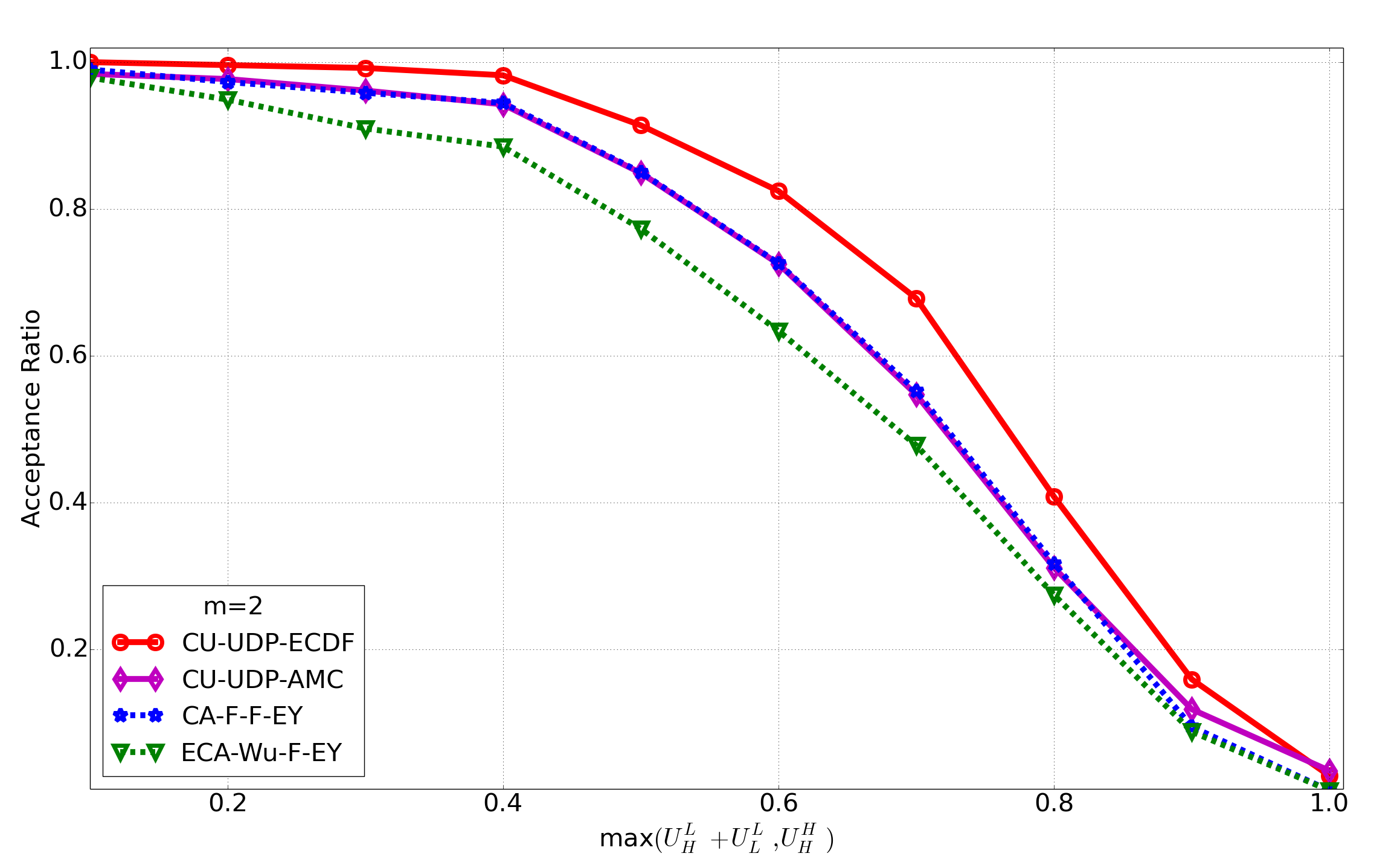}
\caption{$m=2$}
\label{fig:2c}
\end{subfigure}
\begin{subfigure}[b]{0.32\linewidth}
\includegraphics[width=\textwidth,height=4cm]{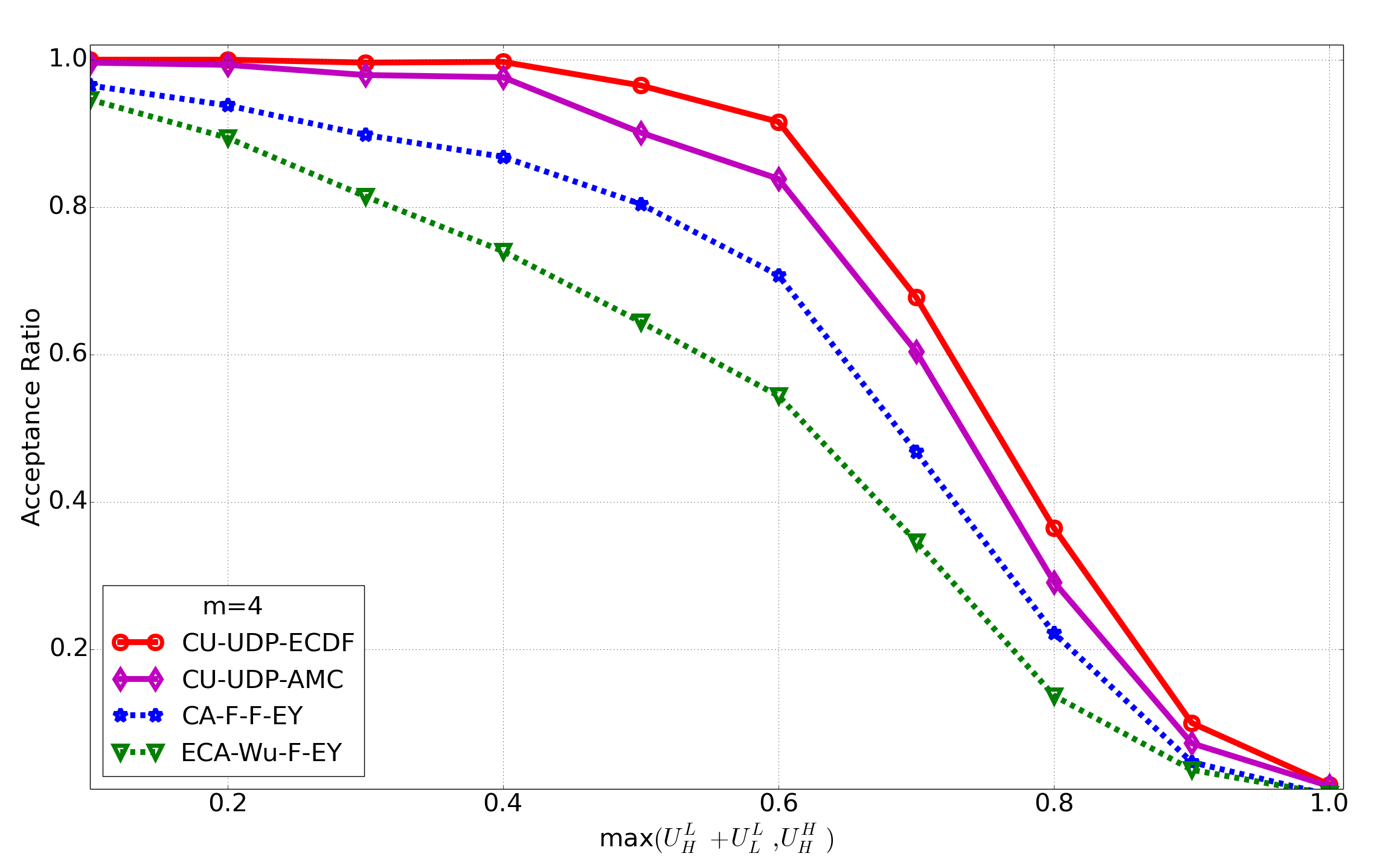}
\caption{$m=4$}
\label{fig:4c}
\end{subfigure}
\begin{subfigure}[b]{0.32\linewidth}
\includegraphics[width=\textwidth,height=4cm]{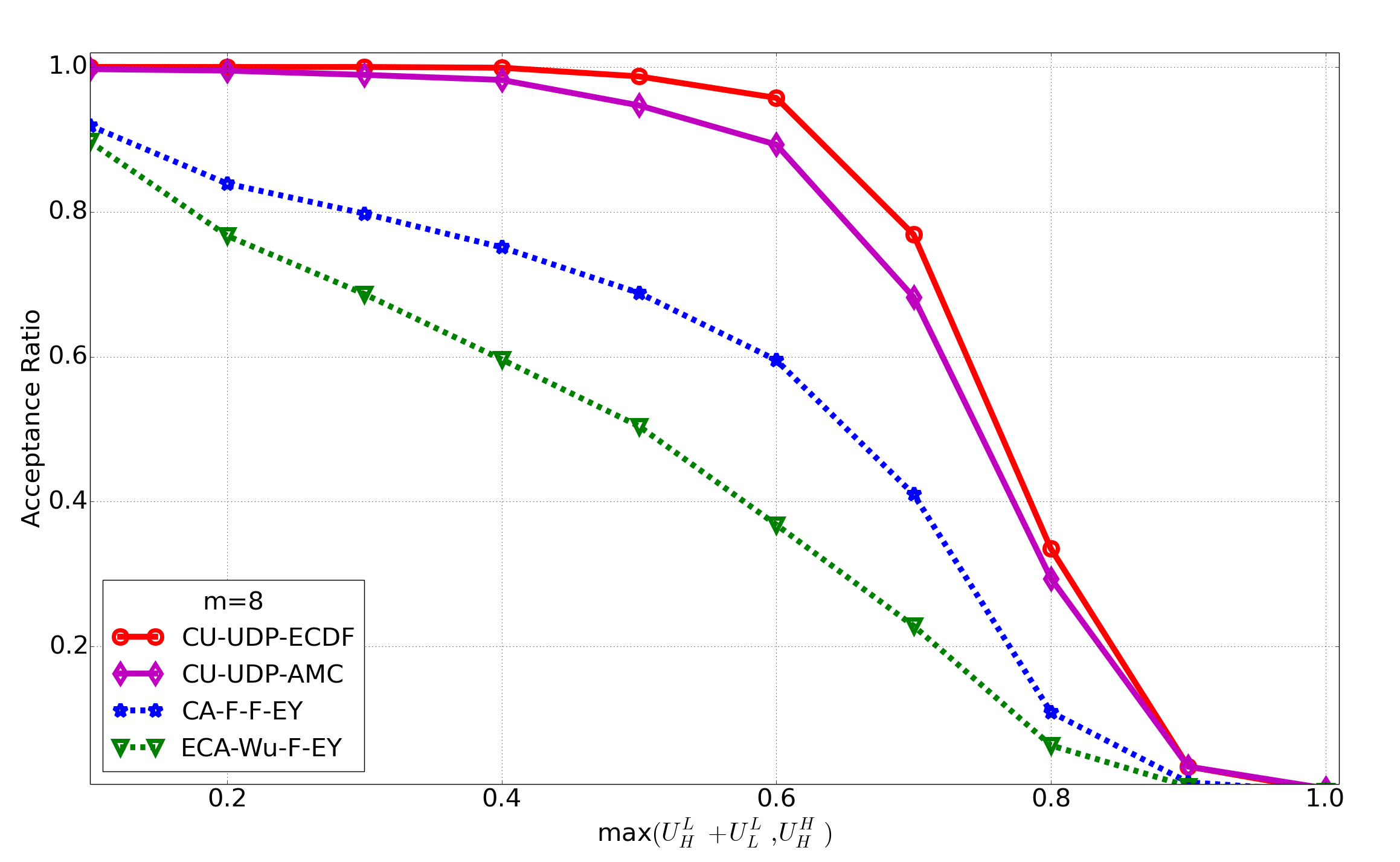}
\caption{$m=8$}
\label{fig:8c}
\end{subfigure}
\caption{Comparison of Acceptance Ratio (constrained-deadline)}\label{schedulability_constrained}
\vspace{-0.2cm}
\end{figure*}
\begin{figure}
\centering
\begin{subfigure}[b]{0.49\textwidth}
\includegraphics[width=0.95\linewidth,height=4cm]{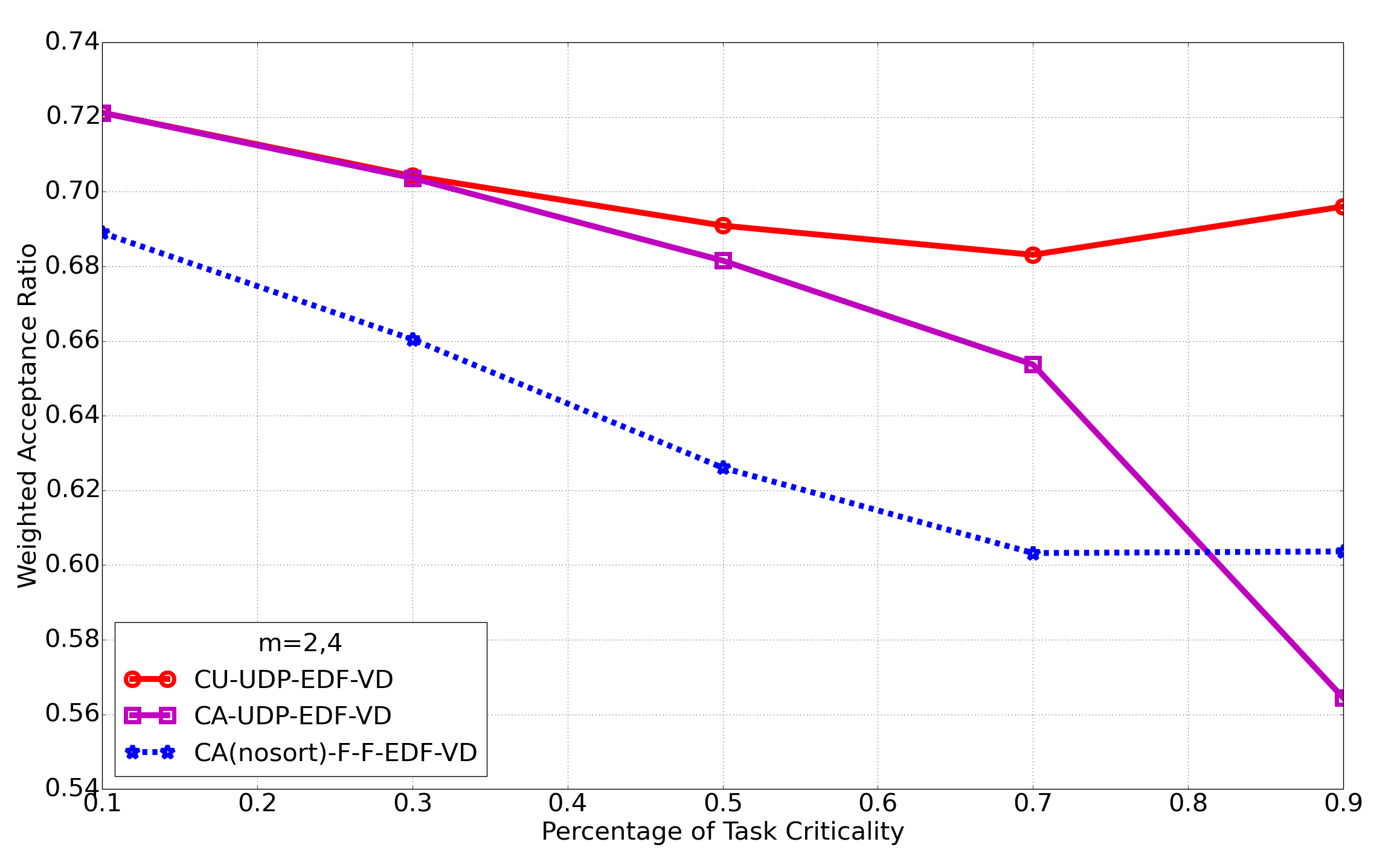}
\caption{Implicit-deadline (EDF-VD)}
\label{fig:war_a}
\end{subfigure}
\begin{subfigure}[b]{0.49\textwidth}
\includegraphics[width=0.95\linewidth,height=4cm]{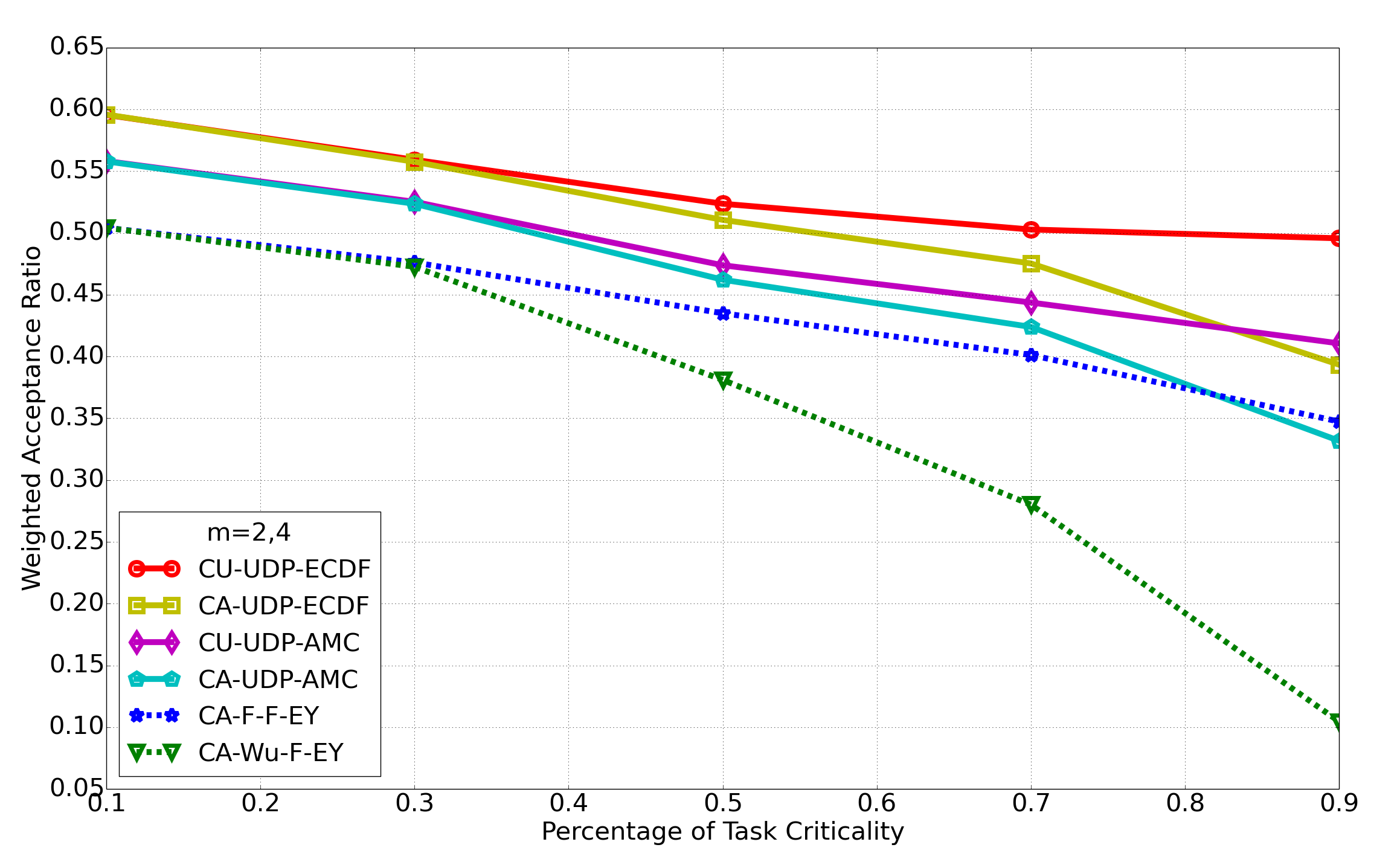}
\caption{Constrained-deadline}
\label{fig:war_c}
\end{subfigure}
\caption{Varying Percentage of Task Criticality}
\label{fig:schedulability_war}
\vspace{-0.1cm}
\end{figure}
\minisection{Experiment Setup} We use the task set generator proposed in~\cite{mcfairgen} for our experiments. The task set parameters used in our generator are as follows:
\begin{itemize}
\item $m \in \{ 2,4,8\}$ is the total number of processors.
\item $u_{min}$ ($= 0.001$) and $u_{max}$ ($= 0.99$) denote the minimum and maximum individual task utilization respectively.
\item The normalized system utilizations are given by $U_H^H$ $\in[0.1, 0.2, \ldots, 0.9, 0.99]$, $U_H^L$ $\in[0.05, 0.15, \ldots, U_H^H]$ and $U_L^L$ $\in[0.05, 0.15, \ldots, 0.99-U_H^L]$.
\item The total number of tasks in a task system is lower bounded by $m+1$ and upper bounded by $5*m$.
\item $P_H=0.5$ denotes the percentage of HC tasks.  
\item The period $T_i$ of a task $\tau_i$ is drawn log-uniformly~(\cite{mrand}) at random from $[10, 500]$. 
\item Utilizations $u_i^L$ and $u_i^H$ are derived using standard techniques~(\cite{mcfairgen,mrand}) that ensure a uniform distribution of values. LC and HC executions $C_i^L$ and $C_i^H$ are then derived as $\lceil u_i^L*T_i\rceil$ and $\lceil u_i^H*T_i\rceil$ respectively.    
\item The deadline $D_i$ of the task $\tau_i$ is drawn uniformly from $[C_i^H, T_i]$ for constrained-deadline task systems.
\end{itemize}

We generate 1000 task sets for each value of total normalized utilization $U_B$ ($U_B=max(U_H^L+U_L^L, U_H^H)$). 

\minisection{Results} In Figures~\ref{schedulability_implicit_speed_up},~\ref{schedulability_implicit} and~\ref{schedulability_constrained}, we present the overall schedulability performance of the algorithms for varying $m$ and $P_H=0.5$, where we plot the acceptance ratios of the algorithms, i.e., fraction of task sets deemed to be schedulable versus total normalized utilization $U_B$. As shown, both CA-UDP and CU-UDP based algorithms perform much better than all the existing algorithms. In addition, the performance gap between our algorithms and the existing algorithms increase as $m$ increases, depicting its scalability.

In Figure~\ref{schedulability_implicit_speed_up}, we compare the performance of UDP strategies with EDF-VD against CA(nosort)-F-F-EDF-VD for implicit-deadline task systems. All the three algorithms use the same EDF-VD schedulability test, and consequently have a speed-up bound of $8/3$. As shown, the performance improvement over CA(nosort)-F-F-EDF-VD due to UDP is as much as $13.3\%$, $22.8\%$ and $28.1\%$ for $m=2$, $4$ and $8$ respectively.

In Figures~\ref{schedulability_implicit} and~\ref{schedulability_constrained}, we compare the performance of UDP strategies with ECDF and AMC against ECA-Wu-F-EY and CA-F-F-EY, for implicit- and constrained-deadline task systems respectively. The performance of CA-UDP based algorithms is similar to that of CU-UDP based algorithms but with the slightly lower performance. For clarity of presentation, we only present the results of CU-UDP based algorithms. 
For implicit-deadline task systems the performance improvement over existing algorithms is as much as $3.2\%$, $3.8\%$ and $9.5\%$ under AMC, and $9.8\%$, $15.2\%$ and $15.7\%$ under ECDF, for $m=2$, $4$ and $8$ respectively. Correspondingly for constrained-deadline task systems, it is as much as $3.5\%$, $13.1\%$ and $29.7\%$ under AMC, and $12.6\%$, $20.8\%$ and $36.2\%$ under ECDF, for $m=2$, $4$ and $8$ respectively.

In Figure~\ref{fig:schedulability_war} we compare the \emph{weighted acceptance ratio} (WAR) of the algorithms for varying $P_H\in\{0.1,0.3,0.5,0.7,0.9\}$ and $m\in\{2,4\}$. Weighted acceptance ratio is defined as $WAR(\mathbb{S}) = \frac{\underset{U_B\in\mathbb{S}}{\sum}(AR(U_B) X U_B)}{\underset{U_B\in\mathbb{S}}{\sum} U_B}$, where $\mathbb{S}$ is the set of $U_B$ values and AR($U_B$) is the acceptance ratio for a specific value of the total normalized utilization $U_B$. Each data point in the plot corresponds to at least $5000$ task sets. 

In Figure~\ref{fig:war_a}, we present the performance of the UDP strategies with EDF-VD for implicit-deadline task systems. As it can be seen, the performance of CA-UDP decreases with increasing percentage of task criticality. At very high $P_H$ values, the task set comprises mainly of HC tasks and as a result, the LC tasks in the system have high utilization values. Hence, CA-UDP performs poorly. In Figure~\ref{fig:war_c}, we compare the performance of UDP strategies with AMC and ECDF for constrained-deadline task systems. The performance of all the strategies with criticality-aware worst-fit allocation of HC tasks decreases as $P_H$ increases. This is expected because when a large number of HC tasks are present in the system, criticality-aware partitioning behaves like a traditional (non-MC) system where worst-fit allocation is known to perform poorly. Algorithms with CU-UDP strategy perform well irrespective of the $P_H$ values.

An interesting observation from these plots is the very good performance of CA-UDP and CU-UDP with AMC. It is worth noting that there is no existing partitioned MC algorithm that employs a fixed-priority scheme such as AMC. This is mainly because AMC is known to perform poorly when compared to dynamic-priority schemes such as ECDF and EY. 
Our results show that in the case of partitioned multiprocessor scheduling, the choice of partitioning strategy has a significant impact on algorithm performance. This improved performance under AMC is also important because fixed-priority scheduling is generally preferred in safety-critical industries.

Another interesting observation is that the performance of CU-UDP strategy is slightly better than CA-UDP, irrespective of the schedulability test being used. This is due to the prioritized allocation of heavy utilization LC tasks, while still ensuring a balanced distribution of the utilization difference of HC tasks under CU-UDP. Although CA-UDP balances this utilization difference, it does so at the cost of non-allocating heavy utilization LC tasks.

Thus, we can conclude that the proposed UDP strategies significantly improve schedulability across a variety of scheduling algorithms. The fact that schedulability under AMC is also better than for strategies using EY, shows that the observed improvements are primarily because of the partitioning strategies, rather than due to difference in schedulability tests. 
\section{Conclusion}
\label{conclusion}

In this paper, we considered the problem of partitioning dual-criticality task systems on a multiprocessor platform. We proposed a utilization difference based partitioning scheme for allocating the HC tasks to the processors, which essentially balances the workload between the two criticality levels on each processor. We compared our partitioning strategies with existing approaches using three different scheduling algorithms. Our results show that the proposed strategies for both criticality-aware and criticality-unaware partitioning perform much better than the existing approaches under both dynamic as well as fixed-priority scheduling.

\section*{Acknowledgement}
This work was funded in part by the Ministry of Education, Singapore, grant number ARC9/14.

\bibliographystyle{IEEEtran}
\bibliography{all}

\end{document}